\documentstyle[12pt]{article}

\makeatletter
\def\vereq#1#2{\lower3pt\vbox{\baselineskip1.5pt \lineskip1.5pt
\ialign{$\m@th#1\hfill##\hfil$\crcr#2\crcr\sim\crcr}}}
\def\lesssim{\mathrel{\mathpalette\vereq<}}
\def\gtrsim{\mathrel{\mathpalette\vereq>}}
\def\alt{\lesssim}
\def\agt{\gtrsim}
\makeatother

\begin{document}

\hfill
\vtop{\hbox{\bf MADPH-00-1187}
      \hbox{\bf VAND-TH-00-7}
      \hbox{\bf AMES-HET 00-11}
      \hbox{August 2000}}

\begin{center}
{\Large\bf Fate of the Sterile Neutrino}\\[4ex]
V. Barger$^1$, B. Kayser$^2$, J. Learned$^3$, T. Weiler$^4$,
and K. Whisnant$^5$\\[2ex]
\it
$^1$Department of Physics, University of Wisconsin, Madison, WI 53706\\
$^2$Division of Physics, National Science Foundation, Arlington, VA 22230\\
$^3$Department of Physics, University of Hawaii, Honolulu, HI 96822\\
$^4$Department of Physics and Astronomy, Vanderbilt University, Nashville, TN
37235\\
$^5$Department of Physics and Astronomy, Iowa State University,
Ames, IA 50011, USA
\end{center}

\vspace{.25in}

\centerline{ABSTRACT}

In light of recent Super-Kamiokande data and global fits that seem to
exclude both pure $\nu_\mu \to \nu_s$ oscillations of atmospheric
neutrinos and pure $\nu_e \to \nu_s$ oscillations of solar neutrinos
(where $\nu_s$ is a sterile neutrino), we reconsider four-neutrino
models to explain the LSND, atmospheric, and solar neutrino oscillation
indications.
We argue that the solar data, with the exception of the $^{37}$Cl
results, are suggestive of $\nu_e \to \nu_s$ oscillations that average
to a probability of approximately ${1\over2}$. In this interpretation,
with two pairs of nearly degenerate mass eigenstates separated by
order $1$~eV, the day-night asymmetry, seasonal dependence, and
energy dependence for $^8$B neutrinos should be small. Alternatively,
we find that four-neutrino models with one mass eigenstate widely
separated from the others (and with small sterile mixings to active
neutrinos) may now be acceptable in light of recently updated LSND
results; the $^{37}$Cl data can be accommodated in this model. For
each scenario, we present simple four-neutrino mixing matrices that
fit the stated criterion and discuss future tests.

\thispagestyle{empty}
\let\Large=\large 

\section{Introduction}

Accelerator, atmospheric, and solar neutrino data give evidence for
neutrino oscillations and thus for neutrino masses and mixing. The LSND
accelerator experiment finds a small $\nu_\mu\to \nu_e$ appearance
probability and a mass-squared difference $\delta m_{\rm LSND}^2 >
0.2\rm\,eV^2$~\cite{LSND,LSND2}. The atmospheric experiments measure the
$\nu_\mu\to\nu_\mu$ survival probability versus both path-length $L$ and
neutrino energy $E_\nu$. The amplitude for $\nu_\mu\to\nu_x$ ($x\neq e$)
oscillations is inferred to be maximal or near-maximal with $\delta
m_{\rm ATM}^2 \sim 3\times
10^{-3}\rm\,eV^2$~\cite{atm,sobel,atm2}. The combined solar
neutrino experiments determine the $\nu_e\to\nu_e$ survival probability
to be $\sim0.3$ to 0.7, depending on the neutrino
energy~\cite{solar,suzuki,solar2,solar3} and $\delta
m_{\rm SOLAR}^2 \alt 10^{-3}\rm\,eV^2$ is required for consistency with
the null result for the $\bar\nu_e\to\bar\nu_e$ probability measurement
in the CHOOZ experiment~\cite{CHOOZ}. Thus, taken together, the data require
three distinct $\delta m^2$. With three neutrinos, there are only two
independent $\delta m^2$, so a fourth sterile neutrino ($\nu_s$) needs
to be invoked in addition to $\nu_e$, $\nu_\mu$, and $\nu_\tau$. Having
no weak interactions, the sterile neutrino escapes the $N_\nu\simeq3$
constraint from the invisible decay width of the $Z$-boson~\cite{PDG}.

Successful global descriptions of the oscillation data have been made in
a four-neutrino
framework~\cite{bilenky,our-4nu,gibbons,mohanty,other4nu}.
Compatibility of the LSND result with null results of
accelerator~\cite{CDHS} and reactor~\cite{BUGEY} disappearance
experiments was found~\cite{bilenky,our-4nu} to favor a $2+2$ mass
spectrum, with the mass-squared differences $\delta m_{\rm SOLAR}^2$ and
$\delta m_{\rm ATM}^2$ separated by the larger difference $\delta m_{\rm
LSND}^2$, over a $1+3$ spectrum, with one mass eigenstate widely
separated from the others. In the simplest of the $2+2$ models, the
oscillations are $\nu_\mu\to\nu_\tau$ (atmospheric) and $\nu_e\to\nu_s$
(solar), or alternatively $\nu_\mu\to\nu_s$ (atmospheric) and
$\nu_e\to\nu_\tau$ (solar). More generally, sterile neutrinos may be
involved in both atmospheric and solar oscillations. In the $2+2$ mixing
schemes, the sterile flavor content must be significant in the solar or
atmospheric oscillations or in both.

New results from Super-Kamiokande (SuperK)~\cite{sobel,suzuki} impact
critically on oscillations to sterile neutrinos.  The zenith angle
dependence of high-energy atmospheric events, along with the neutral
current $\pi^0$ production rate, exclude pure $\nu_\mu\to\nu_s$
oscillations at 99\%~C.L\null. The electron energy dependence of solar
neutrino events, along with the absence of a significant day-night
effect, exclude pure $\nu_e\to\nu_s$ oscillations at 99\%~C.L. when all
of the solar data, including the $^{37}$Cl results, are taken into
account. Thus a dominant involvement of $\nu_s$ in either atmospheric or
solar oscillations is brought into question. Also, LSND has reported new
results~\cite{LSND2} with a somewhat lower average oscillation
probability than before and the KARMEN experiment~\cite{KARMEN} excludes
part of the LSND allowed region. Consequently, previous
exclusions~\cite{bilenky,our-4nu} of the $1+3$ schemes need to be
reassessed. In the context of this new data, we reconsider four-neutrino
oscillation models and the implications for the existence of a sterile
neutrino.

\section{$2+2$ Models}

The $2+2$ models have two nearly degenerate pairs of neutrino mass
eigenstates giving $\delta m_{\rm ATM}^2$ and $\delta m_{\rm SOLAR}^2$
separated by the LSND scale $\delta m_{\rm LSND}^2 \sim 1$~eV$^2$.
Maximal $\nu_3, \nu_2$ mixing describes the atmospheric data and maximal
$\nu_0, \nu_1$ mixing describes the solar data. The mass hierarchy may
be normal ($m_3\gg m_1$) or inverted ($m_3\ll m_1$). The $\delta m_{\rm
SOLAR}^2$ splitting must be smaller than $10^{-3}$~eV$^2$ to avoid the
CHOOZ reactor constraint on $\bar\nu_e\to\bar\nu_e$ oscillations and larger
than the $\sim10^{-10}$~eV$^2$ of just-so vacuum oscillations\cite{just-so} so
that the oscillations average to give an approximately
energy-independent solar $\nu_e$ survival probability.

\subsection{Solar Neutrinos: maximal $\nu_e\to\nu_s$ oscillations}

The new SuperK solar neutrino data show a remarkably flat recoil electron
energy distribution from 5~MeV to 14~MeV, with average value~\cite{suzuki}
\begin{equation}
{\rm data/SSM} = 0.465^{+0.016}_{-0.014} \,,
\label{eq:SuperK}
\end{equation}
where SSM is the Standard Solar Model prediction~\cite{solar3}. The
$^8$B flux normalization in the SSM is somewhat uncertain, and the above
result is suggestive of rapid oscillations with maximal amplitude that
give an average oscillation probability, $\langle P(\nu_e \to \nu_e)
\rangle$, of approximately ${1\over2}$.

Regeneration of $\nu_e$ in the Earth, due to the effects of coherent
forward scattering of the neutrinos with matter~\cite{matter}, causes a
day-night variation in the measured neutrino flux.  The smallness of the
day-night effect $2(D-N)/(D+N) = -0.034 \pm 0.022 \pm
0.013$~\cite{suzuki} in the SuperK data (1.3$\sigma$ from zero) puts
strong constraints on solar solutions with matter
effects~\cite{suzuki,day-night,guth,friedland}. It has been noted that
there can be a
day-night effect even for maximal mixing~\cite{guth,friedland}, but
calculations for active neutrinos show that it is small ($\sim 1$--2~\%)
for $\delta m^2_{\rm solar} \agt 5\times10^{-5}$~eV$^2$ or $\delta m^2_{\rm
solar} \alt 10^{-7}$~eV$^2$~\cite{suzuki}. In $\nu_e\to\nu_s$ oscillations,
the difference in coherent scattering amplitudes in matter is
$\sqrt2 G_F (N_e - N_n/2) \simeq \sqrt2 G_F
N_e/2$ (where $N_e$ and $N_n$ are the electron and neutron number
densities, respectively, and $N_e \sim N_n$ in the Earth). The
corresponding amplitude difference for $\nu_e\to\nu_\mu$ (or
$\nu_e\to\nu_\tau$) is $\sqrt2 G_F N_e$. Therefore, for maximal mixing
the day-night effect should be smaller with a sterile neutrino than with
an active neutrino, but it would not be zero.

The regeneration effect in the earth for neutrinos detected at night
raises the value of $\langle P(\nu_e \to \nu_e) \rangle$ for maximal
mixing with active neutrinos from 0.50 to about 0.54~\cite{guth}. For
sterile neutrinos, we roughly estimate that $\langle P(\nu_e \to \nu_e)
\rangle \simeq 0.52$ in SuperK for maximal mixing with
regeneration. Regeneration also causes a weak dependence on energy of
the solar neutrino suppression. For maximal mixing of $\nu_e$ with an
active neutrino, $\langle P(\nu_e \to \nu_e) \rangle$ is slightly higher
at higher neutrino energies~\cite{guth}, consistent with the weak trend
of the SuperK data; less energy dependence should be present with
$\nu_e$ oscillations to $\nu_s$.

Assuming $\langle P(\nu_e \to \nu_e) \rangle = 0.52$ for maximal $\nu_e
\to \nu_s$ oscillations, the SuperK result in Eq.~(\ref{eq:SuperK})
would be reproduced by a $^8$B flux normalization of $n=0.89$.  The SSM
predictions (in SNU) for the Gallium experiments are (70,~$pp$),
(34,~$^7$Be), (3,~$pep$), (10,~CNO), and (12,~$^8$B), giving a total of
129~SNU; the observed values are $75.4^{+7.8}_{-7.4}$ (SAGE),
$77.5^{+7.5}_{-7.8}$ (GALLEX) and $65.8^{+10.8}_{-10.2}$ (GNO). With the
above $^8$B normalization and $\langle P(\nu_e \to \nu_e) \rangle =
{1\over2}$, the predicted rate in the Gallium experiments is 63.8~SNU.
Thus all of the Gallium measurements are consistent within $2\sigma$ of
the value that would be found for maximal amplitude $\nu_e \to \nu_s$
oscillations that average to ${1\over2}$. The new data from GNO are
especially suggestive of this interpretation. A least-squares fit to the
SuperK and Gallium rates with the $^8$B flux normalization $n$ as a free
parameter gives a best fit value of $n=0.90$ with $\chi^2 = 5.5$ for 3
degrees of freedom, which corresponds to a 14\% goodness of fit. In this
calculation we have not taken into account any regeneration effect for
the Gallium data, which may improve the fit.

The discrepant measurement in the above interpretation is the $^{37}$Cl
value of ${\rm data/SSM} = 0.33\pm0.03$ from the Homestake mine
experiment, which would require a significant energy dependence of the
solar $\nu_e$ flux suppression. Most global fits to neutrino oscillation data
include the $^{37}$Cl data (and often disregard the LSND data) and
then $\nu_e\to\nu_s$ oscillations are excluded. We instead suggest the
possibility that the solar data/SSM flux ratio is relatively flat over
the entire 0.233~MeV to 14~MeV energy range and is described by
$\nu_e\to\nu_s$ oscillations with maximal mixing.

An important test of approximate constant suppression of the solar
neutrino spectrum will be the BOREXINO~\cite{BOREXINO} experiment, which
can measure the $^7$Be component of the solar neutrino flux. Nearly
complete suppression of the $^7$Be component is needed for the $^{37}$Cl
data to be consistent with the suppression of the $^8$B component
measured by SuperK. However, a suppression of the $^7$Be flux similar to
the data/SSM measured by SuperK would favor maximal $\nu_e \to \nu_s$
mixing of solar neutrinos. The SNO~\cite{SNO} and ICARUS~\cite{ICARUS}
experiments will also provide critical tests of the $^8$B flux
suppression, and SNO will test
maximal $\nu_e \to \nu_s$ oscillations through the NC/CC ratio (which
should be the same as the value without oscillations).

\subsection{Mixing matrix for the 2 + 2 model}

We assume a $2+2$ scenario in which one pair of nearly degenerate mass
eigenstates has maximal $\nu_e \to \nu_s$ mixing for solar neutrinos and
the other pair has maximal (or nearly maximal) $\nu_\mu \to \nu_\tau$
oscillations for atmospheric neutrinos. Small off-diagonal mixings
between $\nu_e,\nu_s$ and $\nu_\mu,\nu_\tau$ can
accommodate the LSND results. Neglecting $CP$-violating phases for the
present, an approximate mixing with these properties is
\begin{equation}
\left( \begin{array}{c} \nu_s \\ \nu_e \\ \nu_\mu \\ \nu_\tau \\
\end{array} \right)
\simeq \left( \begin{array}{cccc}
{1\over\sqrt2} & {1\over\sqrt2} & 0 & 0 \\
-{1\over\sqrt2} & {1\over\sqrt2} & \epsilon & \epsilon \\
\epsilon & -\epsilon & {1\over\sqrt2} & {1\over\sqrt2} \\
0 & 0 & -{1\over\sqrt2} & {1\over\sqrt2} \\
\end{array} \right)
\left( \begin{array}{c} \nu_0 \\ \nu_1 \\ \nu_2 \\ \nu_3 \\
\end{array} \right) \,,
\label{eq:U}
\end{equation}
where $|\delta m^2_{10}| \ll |\delta m^2_{32}| \ll |\delta m^2_{21}|$.
In the notation of Ref.~\cite{BDWY}, we have chosen the mixing angles
$\theta_{01} = \theta_{23} = \pi/4$, $\theta_{02} = \theta_{03} = 0$,
and $\theta_{12} = \theta_{13}$, with $\epsilon = \sin\theta_{13}$. The
oscillation probabilities are given in Table~\ref{tab:2+2}. The
parameter $\epsilon$ is determined from the LSND data to be
approximately
\begin{equation}
\epsilon \simeq
\left( {0.016{\rm~eV}^2\over |\delta m^2_{\rm LSND}|} \right)^{0.91} \,,
\label{eq:LSNDamp}
\end{equation}
where $\delta m^2_{\rm LSND}$ is restricted to the range 0.2 to
1.7~eV$^2$ by the BUGEY and KARMEN experiments; for
$\delta m^2_{\rm LSND} \simeq 6$~eV$^2$, $\epsilon \simeq
0.022$ is marginally possible.

Other forms of the mixing matrix (see, e.g.,
Refs.~\cite{bilenky,our-4nu,gibbons,mohanty,other4nu,BDWY}) are also acceptable,
provided that the $\nu_e$--$\nu_s$ mixing is nearly maximal. Future
short and long-baseline experiments will be useful in determining the
mixing matrix. For example, there are no
$\nu_e \to \nu_\tau$ oscillations at short baselines for the mixing in
Eq.~(\ref{eq:U}), but there could be if the $U_{e2}$ and $U_{e3}$ mixing
matrix elements are not equal in magnitude or have different phases
(see, e.g., Refs.~\cite{gibbons,BDWY,ourSBL,donini}), in which case
unitarity requires $U_{\tau0} \ne 0$ and/or $U_{\tau1} \ne 0$. Also, both
$\nu_e \to \nu_\mu$ and $\nu_e \to \nu_\tau$ oscillations are possible
at long baselines, and there can be $CP$ violation in these channels if
the four-neutrino mixing matrix is not real (see, e.g.,
Refs.~\cite{BDWY,donini,ourLBL,LBL}).

\section{$1+3$ Models}

Another way to circumvent the SuperK exclusion of sterile neutrinos is
to assume that the solar and atmospheric oscillations are approximately
described by oscillations of three active neutrinos; then the LSND
result can be explained by a coupling of $\nu_e$ and $\nu_\mu$ through
small mixings with a sterile neutrino that has a mass eigenvalue widely
separated from the others.  This $1+3$ model was previously disfavored
by incompatibility of the LSND result with null results of the CDHS and
BUGEY experiments~\cite{bilenky,our-4nu}.  However, in newly updated
results from the LSND experiment~\cite{LSND2}, the LSND allowed region
is slightly shifted, and this opens a small window for the $1+3$
models. The solar neutrino data can be explained by oscillations of
three active neutrinos in the usual way; the results of the $^{37}$Cl
experiment can be reconciled with those of SuperK in part because
oscillations of $\nu_e$ to active neutrinos in SuperK also show up as
neutral current events (with a rate of about 1/6 of the charge-current
rate), thereby giving a higher data/SSM in SuperK than in the $^{37}$Cl
experiment.


By way of example, consider the approximate mixing
\begin{equation}
\left( \begin{array}{c} \nu_e \\ \nu_\mu \\ \nu_\tau \\ \nu_s \\
\end{array} \right)
\simeq \left( \begin{array}{cccc}
{1\over\sqrt2} & {1\over\sqrt2} & 0 & \epsilon \\
-{1\over2} & {1\over2} & {1\over\sqrt2} & \delta \\
{1\over2} & -{1\over2} & {1\over\sqrt2} & 0 \\
{\delta\over2}-{\epsilon\over\sqrt2}
& -{\delta\over2}-{\epsilon\over\sqrt2}
& -{\delta\over\sqrt2} & 1 \\
\end{array} \right)
\left( \begin{array}{c} \nu_1 \\ \nu_2 \\ \nu_3 \\ \nu_0 \\
\end{array} \right) \,,
\label{eq:U2}
\end{equation}
where $\epsilon$ and $\delta$ are small and both the flavor and mass
eigenstates are reordered to reflect the fact that $\nu_0$ (which
is predominantly $\nu_s$) is the heaviest state. In the notation of
Ref.~\cite{BDWY}, we have chosen the mixing angles $\theta_{01} =
\theta_{12} = \pi/4$, $\theta_{02} = \theta_{23} = 0$, $\sin\theta_{03}
= \epsilon$, and $\sin\theta_{13} = \delta$.
Here the $3\times3$ submatrix that describes the
mixing of the three active neutrinos has the bimaximal
form~\cite{bimax}. The oscillation probabilities are given in
Table~\ref{tab:1+3}. For this mixing matrix the leading oscillation
amplitudes at the $\delta m^2_{\rm LSND}$ scale in the $1+3$ scheme are
(when $\epsilon,\delta \ll 1$)
\begin{equation}
A_{\rm BUGEY} \simeq 4 \epsilon^2
\label{eq:BUGEY}
\end{equation}
for $\bar\nu_e$ disappearance in the BUGEY experiment,
\begin{equation}
A_{\rm CDHS} \simeq 4 \delta^2
\label{eq:CDHS}
\end{equation}
for $\nu_\mu$ disappearance in the CDHS experiment, and
\begin{equation}
A_{\rm LSND} = 4 \epsilon^2 \delta^2
\label{eq:LSND}
\end{equation}
for $\nu_\mu \to \nu_e$ appearance in the LSND experiment. Including
the subleading oscillation at the $\delta m^2_{\rm atm}$ scale, and
assuming that the leading oscillation averages,
\begin{equation}
P(\nu_\mu \to \nu_\mu) \simeq
(1 - 2 \delta^2) (1 - \sin^2 1.27 \delta m^2_{\rm atm} L/E)
\label{eq:atm}
\end{equation}
for $\nu_\mu \to \nu_\mu$ disappearance in atmospheric neutrino
experiments, where $\delta m^2_{\rm atm}$ is in eV$^2$, $L$ in km, and
$E$ in GeV. The relative normalization of the atmospheric $\nu_\mu$ and
$\nu_e$ fluxes is well known~\cite{atmosflux}; the predicted effective change in this normalization at the $\delta m^2_{\rm atm}$ oscillation scale is
$(1 - 2\delta^2)/(1 - 2\epsilon^2)$. In the region where the CDHS
constraint on $\delta$ is weak or nonexistent ($\delta m^2_{\rm LSND}
\alt 0.4$~eV$^2$), BUGEY constrains $\epsilon$ to be less than about
0.1; conservatively assuming that the atmospheric data constrains the
relative normalization to within 10\%, we find $\delta < 0.24$.



%
%

The upper limits on $\epsilon$ (from $A_{\rm BUGEY}$) and $\delta$ (from
$A_{\rm atm}$ and $A_{\rm CDHS}$) and the allowed values of $A_{\rm
LSND}$ vary with $\delta m^2_{\rm LSND}$. With the previously reported
LSND data~\cite{LSND}, there was no value of $\delta m^2_{\rm LSND}$ for
which the constraints on $\epsilon$ and $\delta$ were consistent with
the LSND 99\%~C.L. allowed range of $A_{\rm LSND}$; the maximum allowed
values of $\epsilon$ and $\delta$ always implied an upper limit on
$A_{\rm LSND}$ that was below the LSND measured value.  However, with
the recent shift of $A_{\rm LSND}$ to lower values, there are now three
small $\delta m^2_{\rm LSND}$ islands where the LSND 99\%~C.L. region is
compatible with the BUGEY, CDHS, and KARMEN constraints: $\delta
m^2_{\rm LSND} \simeq 0.9$~eV$^2$ and $\delta m^2_{\rm LSND} \simeq
1.7$~eV$^2$, for which the BUGEY constraint is somewhat less restrictive,
and $\delta m^2_{\rm LSND} \simeq 6$~eV$^2$, for which both the BUGEY and
KARMEN constraints are somewhat less restrictive. The E776 experiment at
BNL~\cite{E776} gives a somewhat tighter constraint than KARMEN at
$\delta m^2_{\rm LSND} = 6$~eV$^2$, but a small region is still allowed
here. Except for a small region near $\delta m^2_{\rm LSND} \simeq
0.2$~eV$^2$ (see below), the combined data are
still inconsistent with oscillations having $\delta m^2_{\rm LSND} <
0.9$~eV$^2$, and for all $\delta m^2_{\rm LSND}$ when the LSND 90\%~C.L.
allowed region is used. The above analysis is summarized in
Table~\ref{tab:constraints}; acceptable values of $\delta m^2_{\rm
LSND}$ are those for which $(4\epsilon^2 \delta^2)_{\rm max}$ lies
within the range of $A_{\rm LSND}$ allowed by LSND and KARMEN.

There is no CDHS constraint for $\delta m^2_{\rm LSND} < 0.25$~eV$^2$,
which suggests that $\delta m^2_{\rm LSND} \simeq 0.2$~eV$^2$ may also
be allowed. However, at these $\delta m^2_{\rm LSND}$ values $\delta$
must be more than 0.5 to reconcile the BUGEY limit with the LSND
measurement, which is not consistent with the relative normalization
of the $\nu_\mu$ and $\nu_e$ fluxes (see above). 

The model of Eq.~(\ref{eq:U2}) has bimaximal mixing in the $3\times3$
active sector. Similar results for the $1+3$ model can be obtained for
any $3\times3$ submatrix that can describe the atmospheric and solar
data for active neutrinos, as long as $\epsilon$ and $\delta$ are small.
This is easily seen by realizing that the
general expressions for the short-baseline amplitudes in the $1+3$
model in Eqs.~(\ref{eq:BUGEY})--(\ref{eq:LSND}) depend only on the
magnitudes of the mixing matrix elements $U_{e0}$ and $U_{\mu0}$
[$\epsilon$ and $\delta$, respectively, in Eq.~(\ref{eq:U2})].
If $U_{\tau0} \ne 0$,
then there can be $\nu_\mu \to \nu_\tau$ and $\nu_e \to \nu_\tau$
oscillations at the $\delta m^2_{\rm LSND}$ scale. If $U_{e3} \ne 0$,
then there can be $\nu_e \to \nu_\mu$ and $\nu_e \to \nu_\tau$
oscillations at the $\delta m^2_{\rm atm}$ scale.

%
%

Because the allowed windows are already severely constrained by
accelerator and reactor data, future experiments at short baselines
could easily test or rule out the $1+3$ scenarios.
MiniBooNE~\cite{miniboone} will search for $\nu_\mu \to \nu_e$
oscillations and test $A_{\rm LSND}$. Reactor experiments such as at
Palo Verde~\cite{PV}, KamLAND~\cite{kamland}, and the proposed
ORLaND~\cite{orland}, will test $A_{\rm BUGEY}$, but only ORLaND is
expected to significantly improve the bound on $\epsilon$.  A future
precision measurement of $\nu_\mu$ disappearance at short baselines
could test $A_{\rm CDHS}$.

\section{Discussion and Conclusions}

The recent Super-Kamiokande data and global fits present new
constraints on the mixing of light sterile neutrinos.    It is
important to assess the viability of a light sterile neutrino, because
its existence is required if the solar, atmospheric, reactor, and
accelerator data are all to be understood in terms of neutrino
oscillations.

We have examined at a qualitative level two schemes for four-neutrino
mass and mixing which nearly accommodate all present data.  In the $2+2$
scheme, two neutrino pairs are separated by the LSND mass scale.  One
pair maximally mixes $\nu_\mu$ and $\nu_\tau$ to explain the atmospheric
data, while the other pair maximally mixes $\nu_e$ and $\nu_s$ to explain
the solar deficit with energy-independent oscillations.  All data except
the low $\nu_e$ capture rate on $^{37}$Cl are explained in this model.
A very small day-night effect is expected, at the level of a per cent or
less. Suppression of the $^7$Be component of the solar neutrino flux
by approximately 50\% is expected for the BOREXINO experiment.

In the $1+3$ scheme, the three active neutrinos are separated by the
LSND scale from a sterile neutrino state. The active neutrino $3\times3$
submatrix explains the atmospheric and solar data, including the
$^{37}$Cl rate. In our example we assigned the bimaximal
model~\cite{bimax} to this submatrix, but any $3\times3$ active
submatrix that describes the atmospheric and solar data may be used.  A small
mixing of $\nu_e$ and $\nu_\mu$ via the sterile state explains the
LSND data.  The smaller oscillation amplitude recently reported by the
LSND collaboration allows marginal accommodation with the null
$\bar\nu_e$ and $\nu_\mu$ disappearance results previously obtained in
reactor and accelerator experiments.

The above two classes of models allow a viable sterile neutrino having
either large mixing with $\nu_e$ or small mixing with all active
neutrinos. In both classes we made the simplifying assumption that
$\nu_\tau$-$\nu_s$ mixing is negligible. Present data may allow
considerable $\nu_\tau$-$\nu_s$ mixing, in which case both solar and
atmospheric neutrino oscillations could have a sizable sterile
component. Quantitative fits to atmospheric and solar data are needed to
determine how much $\nu_\tau$-$\nu_s$ mixing is allowed (see, e..g,
Ref.~\cite{tau-s}).



Neutrinos also provide a hot dark matter component, which is relevant to
large-scale structure formation~\cite{primack}. {}From the upper limit
of $m_\beta = 2.5$ to 3~eV on the effective $\nu_e$ mass obtained from tritium
beta decay endpoint measurements\cite{lobashev}, the mass $m_{\rm max}$ of the heaviest
neutrino is bounded by~\cite{BWW}
\begin{equation}
\sqrt{\delta m_{\rm LSND}^2} \alt m_{\rm max}
\alt \sqrt{m_\beta^2 + \delta m_{\rm LSND}^2} \,.
\end{equation}
The contribution of the neutrinos to the mass density of the universe is
given by $\Omega_\nu = \sum m_\nu/(h^2 93$~eV), where $h$ is the present
Hubble expansion parameter in units of 100 km/s/Mpc \cite{expansion};
with $h=0.65$, both the $2+2$ and $1+3$ models give $\Omega_\nu \ge
0.02$. For an inverted neutrino mass spectrum (where the $\nu_e$ is
associated predominantly with the heavier neutrinos), the bound is
$\sqrt{\delta m^2_{\rm LSND}} \alt m_{\rm max} \alt m_\beta$, which
yields $\Omega_\nu \ge 0.02$ and $\Omega_\nu \ge 0.07$ for the $2+2$ and
$1+3$ models, respectively. The MAP\cite{MAP} and PLANCK\cite{PLANCK}
satellite measurements of the cosmic microwave background radiation may
be sensitive to these neutrino densities.

The existence of the LSND mass gap will be tested by the MiniBooNE
experiment~\cite{miniboone}.  A definitive test of the
presence of the sterile state in the $2+2$ model will be the measurement
of the NC/CC ratio for solar neutrinos by the SNO experiment.  With the
solar solution given by maximal $\nu_e \to \nu_s$ mixing in this model,
the NC should show the same suppression as the CC.  The existence of the
sterile neutrino in the $1+3$ model will not be tested by SNO NC/CC data since
the ratio is approximately that of $\nu_e$ oscillations to active
neutrinos. The sterile state in the $1+3$ model can be tested by
searches for small amplitude oscillations at short baselines.

Observation of the flavor ratio of extragalactic
neutrinos may serve as a test of the models.  Oscillations of
neutrinos from distant sources will have averaged, leaving definite
predictions for flavor ratios.  If cosmic neutrinos are mainly produced
in pion/muon decay, their initial flavor ratio is $\nu_\tau :\nu_\mu:
\nu_e :\nu_s \approx 0:2:1:0$.  A simple calculation~\cite{LPetc} then
gives the asymptotic ratios $1:1:0.5:0.5$ for the $2+2$ model with
two maximally-mixed pairs, and $1:1:1:0$ for the $1+3$ model with
bimaximal mixing of the active neutrinos.

A virtue of active-sterile neutrino oscillations is that they may aid
r-process nucleosynthesis of heavy elements in neutrino-driven
supernovae ejecta.  The basic requirement is that the $\nu_e$ flux be
diminished in the region where inverse $\beta$-decay would otherwise
transform neutrons into protons.  The $2+2$ and $1+3$ mass spectra and
mixing matrices presented here are of the forms previously discussed to
enhance r-process nucleosynthesis.  In the $2+2$ model, this is
accomplished with a two-step process: first, a matter-enhanced
$\nu_\mu/\nu_\tau \to \nu_s$ transition beyond
the neutrino sphere removes the energetic $\nu_\mu/\nu_\tau$ before they
can convert to energetic $\nu_e$, and then a large
$\nu_e \to \nu_\mu/\nu_\tau$ transition reduces the $\nu_e$
abundance~\cite{twoPtwo}; the required mass-squared parameter is $\delta
m^2_{\rm LSND} \agt 1 {\rm~eV}^2$.  In the $1+3$ scheme, the relevant
oscillation is a matter-enhanced $\nu_e \to \nu_s$, obtained with
$\delta m^2_{\rm LSND} \agt 2 {\rm~eV}^2$ and $\sin^2 2\theta_{es}\agt
10^{-4}$~\cite{threePone}. It appears that the r-process
enhancements discussed in Refs.~\cite{twoPtwo,threePone} may be
accomplished by tuning the small parameters in our mixing schemes.


\bigskip
\noindent
{\bf Acknowledgments}: We appreciate the hospitality of the Aspen Center for Physics, where this work was initiated. We thank Matt Lautenschlager and Ben Wood for useful interactions.
This research was supported in part by the U.S.~Department of Energy
under Grants No.~DE-FG02-95ER40896, No.~DE-FG05-85ER40226, and
No.~DE-FG02-94ER40817, and in part
by the University of Wisconsin Research Committee with funds granted
by the Wisconsin Alumni Research Foundation.

\newpage

\begin{table}[h]
\caption[]{Oscillation probabilities in the $2+2$ model defined by
Eq.~(\ref{eq:U}), to leading order for each oscillation scale. The
oscillation arguments are defined by $\Delta_j \equiv 1.27 \delta m^2_j
L/E$, with $\delta m^2_j$ in eV$^2$, $L$ in km, and $E$ in GeV.
\label{tab:2+2}}
\centering\leavevmode
\begin{tabular}{|l|c|c|c|c|}
\hline\hline
 & $s$ & $e$ & $\mu$ & $\tau$\\
\hline
$s$& $1 -  \sin^2\Delta_{\rm SOLAR}$&
$\sin^2\Delta_{\rm SOLAR}$& \ $2\epsilon^2 \sin^2\Delta_{\rm SOLAR}$& 0\\
\hline
$e$&  &
$\begin{array}{c}
    1 - 8\epsilon^2 \sin^2\Delta_{\rm LSND}\\
    {} -  \sin^2\Delta_{\rm SOLAR}
\end{array}$&
$\begin{array}{c}
    8\epsilon^2\sin^2\Delta_{\rm LSND}\\
    {} - 2\epsilon^2\sin^2\Delta_{\rm ATM}\quad\ \\
    {} - 2\epsilon^2\sin^2\Delta_{\rm SOLAR}\ \
\end{array}$&
$2\epsilon^2\sin^2\Delta_{\rm ATM}$\\
\hline
$\mu$&  &  &
$\begin{array}{c}
    1 - 8\epsilon^2\sin^2\Delta_{\rm LSND}\\
    {} - \sin^2\Delta_{\rm ATM}
\end{array}$&
$\sin^2\Delta_{\rm ATM}$\\
\hline
$\tau$& & & & $1 - \sin^2\Delta_{\rm ATM}$\\
\hline\hline
\end{tabular}
\end{table}

\begin{table}[h]
\caption[]{Oscillation probabilities in the $1+3$ model defined by
Eq.~(\ref{eq:U2}), to leading order for each oscillation scale. The
oscillation arguments are defined by $\Delta_j \equiv 1.27 \delta m^2_j
L/E$, with $\delta m^2_j$ in eV$^2$, $L$ in km, and $E$ in GeV.
\label{tab:1+3}}
\centering\leavevmode
\begin{tabular}{|l|c|c|c|c|}
\hline\hline
& $e$ & $\mu$ & $\tau$ & $s$\\
\hline
$e$
& $\begin{array}{c}
    1 - 4\epsilon^2 \sin^2\Delta_{\rm LSND}\\
    {} -  \sin^2\Delta_{\rm SOLAR}
\end{array}$
& $\begin{array}{c}
    {} 4\epsilon^2\delta^2\sin^2\Delta_{\rm LSND}\\
    + {1\over2}\sin^2\Delta_{\rm SOLAR}
\end{array}$
& ${1\over2}\sin^2\Delta_{\rm SOLAR}$
& $\begin{array}{c}
    {} 4\epsilon^2\sin^2\Delta_{\rm LSND}\\
    + {1\over2}(\delta^2-2\epsilon^2)\sin^2\Delta_{\rm SOLAR}
\end{array}$\\
\hline
$\mu$ &
& $\begin{array}{c}
    1 - 4\delta^2 \sin^2\Delta_{\rm LSND}\\
    {} -  \sin^2\Delta_{\rm ATM}\\
    {} -  {1\over4}\sin^2\Delta_{\rm SOLAR}
\end{array}$
& $\begin{array}{c}
    {}\sin^2\Delta_{\rm ATM}\\
    - {1\over4}\sin^2\Delta_{\rm SOLAR}
\end{array}$
& $\begin{array}{c}
    {} 4\delta^2 \sin^2\Delta_{\rm LSND}\\
    -  \delta^2\sin^2\Delta_{\rm ATM}\\
    +  {1\over4}(2\epsilon^2-\delta^2)\sin^2\Delta_{\rm SOLAR}
\end{array}$
\\
\hline
$\tau$ &  &
& $\begin{array}{c}
    1 - \sin^2\Delta_{\rm ATM}\\
    {} - {1\over4}\sin^2\Delta_{\rm SOLAR}
\end{array}$
& $\begin{array}{c}
    {} \delta^2\sin^2\Delta_{\rm ATM}\\
    + {1\over4}(2\epsilon^2-\delta^2)\sin^2\Delta_{\rm SOLAR}
\end{array}$
\\
\hline
$s$ & & &
& $\begin{array}{c}
1 - 4(\epsilon^2+\delta^2)\sin^2\Delta_{\rm LSND}\\
{} - \delta^2(2\epsilon^2+\delta^2)\sin^2\Delta_{\rm ATM}
\end{array}$
\\
\hline\hline
\end{tabular}
\end{table}

\begin{table}[h]
\caption[]{Summary of constraints on four-neutrino oscillation parameters
in the 1+3 scheme. The upper limit on $A_{\rm LSND}$ at $\delta m^2_{\rm
LSND} \simeq 6$~eV$^2$ is from the E776 experiment at BNL~\cite{E776}.
All experimental limits are at 90\%~C.L., except
for LSND, which is at 99\%~C.L. \label{tab:constraints}}
\centering\leavevmode
\begin{tabular}{|l|c|c|c|c|c|c|}
\hline\hline
$\delta m^2_{\rm LSND}$ & $\epsilon_{\rm max}$
& $\delta_{\rm max}$
& $(4\epsilon^2\delta^2)_{\rm max}$
& $(A_{\rm LSND})_{\rm min}$
& $(A_{\rm LSND})_{\rm max}$\\
(eV$^2$) & (BUGEY) & (CDHS) & & (LSND) & (KARMEN)\\
\hline
6.0 & 0.19 & 0.14 & $2.8\times10^{-3}$
& $1.5\times10^{-3}$ & $2.0\times10^{-3}$ \\
1.7 & 0.16 & 0.12 & $1.5\times10^{-3}$
& $0.8\times10^{-3}$ & $1.0\times10^{-3}$ \\
0.9 & 0.12 & 0.16 & $1.5\times10^{-3}$
& $1.4\times10^{-3}$ & $3.0\times10^{-3}$ \\
0.3 & 0.10 & 0.45 & $8\times10^{-3}$
& $10\times10^{-3}$ & $30\times10^{-3}$ \\
\hline\hline
\end{tabular}
\end{table}

\end{document}